\documentclass[floatfix,lengthcheck,showpacs,amssymb,amsmath,amsfonts,twocolumn]{aastex63}

\usepackage{lineno}
\usepackage[utf8]{inputenc}
\usepackage{color}
\usepackage{graphicx}
\usepackage{float}
\usepackage{subfigure}
\usepackage{amsmath}
\usepackage{afterpage}
\usepackage{acronym}
\usepackage{multirow}
\usepackage{tikz}
\usetikzlibrary{arrows,shapes,trees,decorations.pathreplacing}
\usepackage{import}
\usepackage{svn-multi}
\usepackage{lineno}
\usepackage{hyperref}
\usepackage{gensymb}
\hypersetup{
    colorlinks=true,
    linkcolor=blue,
    filecolor=magenta,      
    urlcolor=cyan,
}

\newcommand{\msun}{\ensuremath{\mathrm{M}_{\odot}}}

\begin{document}
\title[]{Gravitational-wave Merger Forecasting: Scenarios for the early detection and localization of compact-binary mergers with ground based observatories}

\correspondingauthor{Alexander H. Nitz}
\email{alex.nitz@aei.mpg.de}

\author[0000-0002-1850-4587]{Alexander H. Nitz}
\affil{Max-Planck-Institut f{\"u}r Gravitationsphysik (Albert-Einstein-Institut), D-30167 Hannover, Germany}
\affil{Leibniz Universit{\"a}t Hannover, D-30167 Hannover, Germany}
\author[0000-0002-6990-0627]{Marlin Schäfer}
\affil{Max-Planck-Institut f{\"u}r Gravitationsphysik (Albert-Einstein-Institut), D-30167 Hannover, Germany}
\affil{Leibniz Universit{\"a}t Hannover, D-30167 Hannover, Germany}
\author[0000-0001-5078-9044]{Tito Dal Canton}
\affil{Universit{\'e} Paris-Saclay, CNRS/IN2P3, IJCLab, 91405 Orsay, France}


\begin{abstract}
We present the prospects for the early (pre-merger) detection and localization of compact-binary coalescences using gravitational waves over the next 10 years. Early warning can enable the direct observation of the prompt and early electromagnetic emission of a neutron star merger. We examine the capabilities of the ground based detectors at their ``Design'' sensitivity (2021-2022), the planned ``A+'' upgrade (2024-2026), and the envisioned ``Voyager'' concept (late 2020's).  We find that for a fiducial rate of binary neutron star mergers of $1000 ~\mathrm{Gpc}^{-3} \mathrm{yr}^{-1}$, the Design, A+, and Voyager era networks can provide 18, 54, and 195s of warning for one source per year of observing, respectively, with a sky localization area $<$100 deg$^2$ at a $90\%$ credible level. At the same rate, the A+ and Voyager era networks will be able to provide 9 and 43s of warning, respectively, for a source with $<$10 deg$^2$ localization area. We compare the idealized search sensitivity to that achieved by the PyCBC Live search tuned for pre-merger detection. The gravitational-wave community will be prepared to produce pre-merger alerts. Our results motivate the operation of observatories with wide fields-of-view, automation, and the capability for fast slewing to observe simultaneously with the gravitational-wave network. 

\end{abstract}

\section{Introduction}

The second generation of gravitational-wave observation began in 2015 with the operation of the twin LIGO observatories~\citep{TheLIGOScientific:2014jea}. During the first observing run, the first binary black hole mergers were detected, which provided insight into gravity in the strong-field regime~\citep{abbott:2016blz, GW150914:TGR,TheLIGOScientific:2016pea}. The true era of gravitational-wave multi-messenger astronomy, however, began with GW170817, the first observation of a binary neutron star (BNS) merger with gravitational waves~\citep{TheLIGOScientific:2017qsa}. Only a couple seconds following the gravitational-wave signal observed by the LIGO and Virgo~\citep{TheVirgo:2014hva} observatories, a gamma-ray burst was observed by Fermi-GBM and INTEGRAL~\citep{Goldstein:2017mmi,Savchenko:2017ffs,Monitor:2017mdv}.
About 11 hours later, the optical counterpart was spotted~\citep{Coulter:2017wya} and to date GW170817 has been observed by over 70 observatories spanning the electromagnetic band and including neutrino and cosmic-ray observatories (see ~\cite{GBM:2017lvd} and references therein for a detailed summary). The observation of GW170817 has provided an unprecedented look into the nuclear equation of state~\citep{properties_gw170817, Abbott:2018exr, Radice_2018, Kiuchi_2019, Capano:2019eae, Abbott_2020}, the Hubble constant~\citep{Guidorzi:2017ogy,Hotokezaka:2018dfi,Fishbach:2018gjp}, the phenomenon of kilonova (see ~\cite{Metzger:2019zeh} and references therein), and the central engine of short gamma-ray bursts~\citep{Murguia-Berthier:2020tfs,Wu:2019rla,Lazzati:2020vbo}.

However, a crucial gap in these observations are the records of the early time behavior of the optical emission. Optical observations only began hours after the neutron star merger due to the presence of non-Gaussian transient noise in the LIGO-Livingston data requiring manual intervention and preventing the initial automated release of a precise sky localization~\citep{TheLIGOScientific:2017qsa}. Earlier optical and in particular ultraviolet observations would have been able to differentiate kilonova emission models~\citep{Arcavi:2018mzm}. While the typical latency for automated gravitational-wave alerts has been reduced over time to between tens of seconds to minutes after merger~\citep{LIGOScientific:2019gag}, the holy grail would be to observe a coalescence's prompt and early electromagnetic emission just a matter of seconds after merger. There may be electromagnetic emission which occurs before the merger~\citep{Hansen:2000am, Troja_2010, TsangShattering, Metzger:2016mqu, Wang:2016dgs, Wada:2020kha}. A broad summary of the scientific potential of neutron star merger observations can be found in~\cite{Burns:2019tqz}.

To date, the LIGO and Virgo observatories have detected dozens of gravitational wave sources~\citep{Nitz:2018imz,Nitz:2019hdf,Nitz:2020naa,Venumadhav:2019tad,Venumadhav:2019lyq,Zackay:2019btq,LIGOScientific:2018mvr, GW190412, GW190814, GW190521}, two BNS mergers~\citep{TheLIGOScientific:2017qsa,Abbott:2020uma}, but only a single source, GW170817, had clear electromagnetic counterparts~\citep{GBM:2017lvd,Nitz:2019bxt}.  During O3 there was an active follow-up campaign involving numerous telescopes (see e.g. follow-up of GW190425~\citep{GCN190425}), which included but was not limited to follow-up by Swift, ZTF~\cite{Anand:2020eyg}, MASTER, and GRANDMA~\cite{Antier:2020nuy}. However, over the coming decade, we expect the sensitivity and capability of ground based gravitational-wave observatories to dramatically increase~\citep{Aasi:2013wya}. To match this, the infrastructure for both the low-latency~\citep{Messick:2016aqy,Adams:2015ulm, Hooper:2011rb,Klimenko:2015ypf} and pre-merger detection of gravitational waves is being actively developed by multiple groups~\citep{Cannon:2011vi,Chu:2015jxa, Kapadia:2020kss, Sachdev:2020lfd} with a preliminary test recently conducted after the end of the third observing run (O3)~\citep{O3premergertest}. To take advantage of advance warning, facilities will need automated operation, wide effective fields-of-view, and the ability to rapidly point.

In this letter, we explore the increasing capability of the global gravitational-wave network to detect inspiralling binaries seconds to minutes before merger. We examine the distribution of detectable sources and the evolution of their distance and sky localization over the next several years. Finally, we adapt the existing low-latency search PyCBC Live~\citep{Nitz:2018rgo,DalCanton:2020vpm} to gauge if the current search methods will continue to be suited for pre-merger detection with the forthcoming global network.

\section{Pre-merger detection of mergers with gravitational waves}
\label{sec:sims}

Orbiting compact binaries emit gravitational waves and, due to the loss of orbital energy, inspiral and eventually merge~\citep{Peters:1964}. For low mass sources, such as BNSs, this `inspiral' phase of the gravitational wave signal is the observable portion. The merger and post-merger gravitational-wave signals are buried in the noise for current instruments, as they occur at frequencies ($\sim$1-4 kHz) beyond the detector's most sensitive band~\citep{Clark:2015zxa}.

The most sensitive methods for the detection of compact binary mergers use matched filtering, along with knowledge of a source's expected gravitational waveform~\citep{findchirp}. The waveform model is typically derived from the post-Newtonian expansion of general relativity for neutron star binaries~\citep{Wagoner:1976am, Blanchet:1989ki, Blanchet:2014}. This procedure is optimal in Gaussian noise to detect a signal from a source with known parameters. The gravitational waveform encodes the properties of the source binary (such as the components' masses and spins) on the frequency evolution. To cover a broad region of the unknown source parameters, a discrete set of template waveforms, each representing a possible combination of source parameters, is searched. 

Analyses differ in the exact procedure~\citep{findchirp, Usman:2015kfa, gstlal-methods, Venumadhav:2019tad}, but conceptually, all model-based searches for gravitational-wave signals use such a template bank along with matched filtering to extract the signal-to-noise ratio (SNR) produced by the source in the data. Possible candidates are identified, and their SNR is combined across multiple detectors and statistically assessed. Current low-latency analyses typically take 5-30~s from initial data collection to the final assessment and identification of a candidate~\citep{Nitz:2018rgo, DalCanton:2020vpm, Messick:2016aqy, Hooper:2011rb}.

With this procedure in mind, the goal of pre-merger detection is to identify a candidate gravitational-wave signal early enough to produce an alert before the actual merger is observable from Earth. This means that an initial assessment of the candidate must be made with only the early time (or correspondingly the low frequency) portion of the gravitational-wave signal. To search for the signal at different times before merger, one can expand the idea of the template bank to include a discretization over time before merger. For a particular time before merger, one models the gravitational-wave signal up to only that point in time, and conducts a search in an identical manner as standard analyses. In this way, an identification can be made before the data at time of merger is even collected.

\begin{figure}[!htb]
    \centering
    \includegraphics[width=\columnwidth]{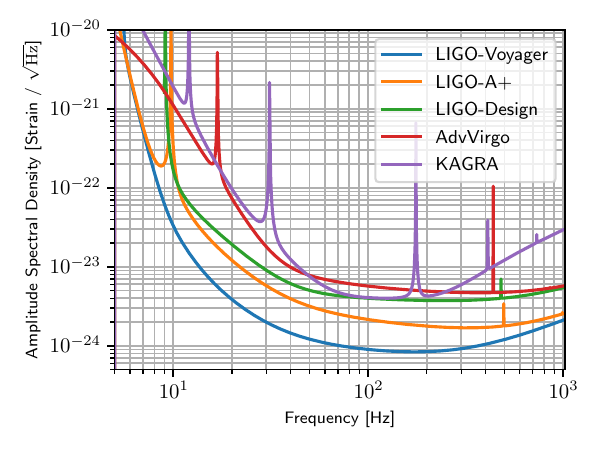}
    \caption{Noise curves for the Advanced Virgo (red), KAGRA (purple), and LIGO instruments used in this study. The LIGO instruments are expected to achieve the ``Design'' curve (green) in the early 2020's, followed by the ``A+'' (orange) in 2024-2026, and ``Voyager'' (blue) in the late 2020's.}
\label{fig:asd}
\end{figure}

\subsection{Observatories and simulated source population}

In this study, we consider the five ground based observatories currently in operation or under construction, namely LIGO-Hanford (H), LIGO-Livingston (L)~\citep{TheLIGOScientific:2014jea}, LIGO-India (I)~\citep{M1100296}, Virgo (V)~\citep{TheVirgo:2014hva}, and KAGRA (K)~\citep{Akutsu:2018axf}. We split our analysis into three sensitivity epochs in different configurations. We denote these epochs as the ``Design'' era which covers the expected sensitivity and operation of the detector network starting from 2021-2022, the ``A+'' era which is timed for the next planned upgrade to the LIGO instruments expected to begin operation in 2024-2026, with LIGO-India joining with equivalent sensitivity towards towards the end of this period~\citep{Aasi:2013wya}, and finally the ``Voyager'' era which includes proposed upgrades to the LIGO instruments predicted to begin operation in the late 2020's~\citep{voyager}.

To assess each detector network, we produce a simulated population of $O(10^5)$ BNS mergers, which are uniformly distributed in volume, and isotropic in binary orientation and sky location. For simplicity of comparison, we choose a reference binary with component masses $1.4-1.4~\msun$. However, as described in Sec.~\ref{sec:scale}, our results can be applied to a more generic population. The gravitational waveform is calculated using TaylorF2, a model based on the post-Newtonian approximation to GR~\citep{Sathyaprakash:1991mt,Droz:1999qx,Blanchet:2002av,Faye:2012we} which is suitable for long duration signals where the merger happens at $\gtrsim 500$~Hz. Each simulated source is added to Gaussian noise colored with the power spectral density corresponding to each instrument at a particular epoch.

There is significant uncertainty in the actual noise capability that will be achieved by each instrument over time. For the LIGO-Hanford and Livingston detectors, we use the ``Design'', ``A+'', and ``Voyager'' noise curves consistent with~\cite{Aasi:2013wya,aplusnoise,voyagernoise}. As done in~\cite{Aasi:2013wya}, we assume LIGO-India will join the network in the mid 2020s using the ``A+'' configuration, and from then on will match the sensitivity of the other LIGO observatories. Note, that we use the Virgo design curve in all cases, consistent with the conservative projection from the mid 2020's in~\cite{Aasi:2013wya}. For KAGRA, we use its design curve~\citep{kagranoise}. Future upgrades to Virgo and KAGRA in the late 2020's may increase their sensitivity during the Voyager era beyond what we consider here. A comparison of these noise curves is shown in Fig.~\ref{fig:asd}.

The advance warning capabilities of the network will depend on the instruments meeting their sensitivity targets at low frequencies. For example, if the ``Voyager'' era instruments only match the sensitivity of the ``A+'' instruments below a gravitational-wave frequency of 30 Hz (a sensitivity reduction of $\sim2-2.5\times$ in this band), then the detection and localization capabilities at times earlier than $\sim 60$s before merger will only match those predicted for the ``A+'' era. Closer to merger, as more of the signal-to-noise is accumulated from higher frequencies, the results would converge to those we show in the next section, assuming the predicted high frequency sensitivity is obtained.

\subsection{Source detection and localization}

We consider two criteria to define whether a particular simulated source is detected at a given time before merger, and hence measure the capabilities of future detector networks.

The first criterion is an idealized, simplified analysis which detects any signal having a total network SNR $>10$. This choice is consistent with the threshold for confidently detected mergers in \cite{LIGOScientific:2018mvr,Nitz:2019hdf}. In practice, we may expect a marginally lower threshold (i.e.~higher sensitivity), dependent on the rate of confounding non-Gaussian noise transients in future detector networks. For nearly Gaussian data, a threshold of $\sim9$ would increase the overall detection rate by $\sim30-40\%$, however, the impact for well-localized sources would be less pronounced. For each simulated signal, we calculate its network SNR as a function of time before merger. After a source reaches the required SNR threshold, we generate the posterior distribution for its spatial localization at each time step before merger. The localization is performed via the rapid Bayestar algorithm, commonly used in production low-latency analyses~\citep{Singer:2015ema}.

The second detection criterion involves an actual analysis of the simulated data with PyCBC Live~\citep{Nitz:2018rgo,DalCanton:2020vpm}, based on the open-source PyCBC gravitational-wave data analysis library~\citep{pycbc-github}.  PyCBC Live is one of several low-latency analyses~\citep{Messick:2016aqy,Adams:2015ulm, Hooper:2011rb,Klimenko:2015ypf} currently used for the rapid detection of gravitational waves by the LIGO-Virgo-KAGRA scientific collaboration and has already been instrumental in the analysis and detection of numerous sources since the second observing run of second-generation detectors~\citep{LIGOScientific:2019gag}. PyCBC Live is computationally efficient, supports searching using arbitrary number of detectors, and is easily reconfigurable, making it suitable for our analysis. Comparing the results of PyCBC Live with our first detection criterion allows us to establish the performance of current low-latency analyses with respect to future detector networks.

PyCBC Live can search for pre-merger signals by using a template bank of truncated TaylorF2 waveforms that discretely sample the time before merger. As the frequency evolution of a TaylorF2 inspiral is monotonic in time, we can truncate it at a particular frequency to approximate a waveform that is similarly truncated at a chosen time before merger. For our analysis, we choose the frequencies corresponding to increments of $5\%$ of the total expected SNR. For each frequency cutoff, we generate a template bank using a standard geometric placement algorithm~\citep{Brown:2012qf}.

We configure PyCBC Live to minimize the latency incurred by the analysis. By careful choice of the analysis step size, power spectral density estimation, and data preconditioning, we reduce the latency of the PyCBC Live analysis down to a worst case delay of $2.5$~s (average of $2$~s), measured from the time data is available to the analysis to the moment a candidate is identified. The latency includes the need to collect data for filtering, the computational processing, and also the latency incurred due to the discreteness of the input data ($1$~s).

In addition to the latency introduced by the search, an extra latency of at least a few seconds was typically introduced by the other steps of the alert generation during O3: the calibration of the strain data ($\sim 3$ s; \cite{VietsThesis}) and its distribution to the computing center, the rapid sky localization ($\sim 1$ s; \cite{BAYESTARSpeedup}), and the processing and public distribution of the alert. Hence, significant work across all steps will be needed to reduce the total latency below $\sim 5$--$10$ s.

\section{Detection and Localization Capabilities}

\begin{figure*}
    \centering
    \includegraphics[width=\textwidth]{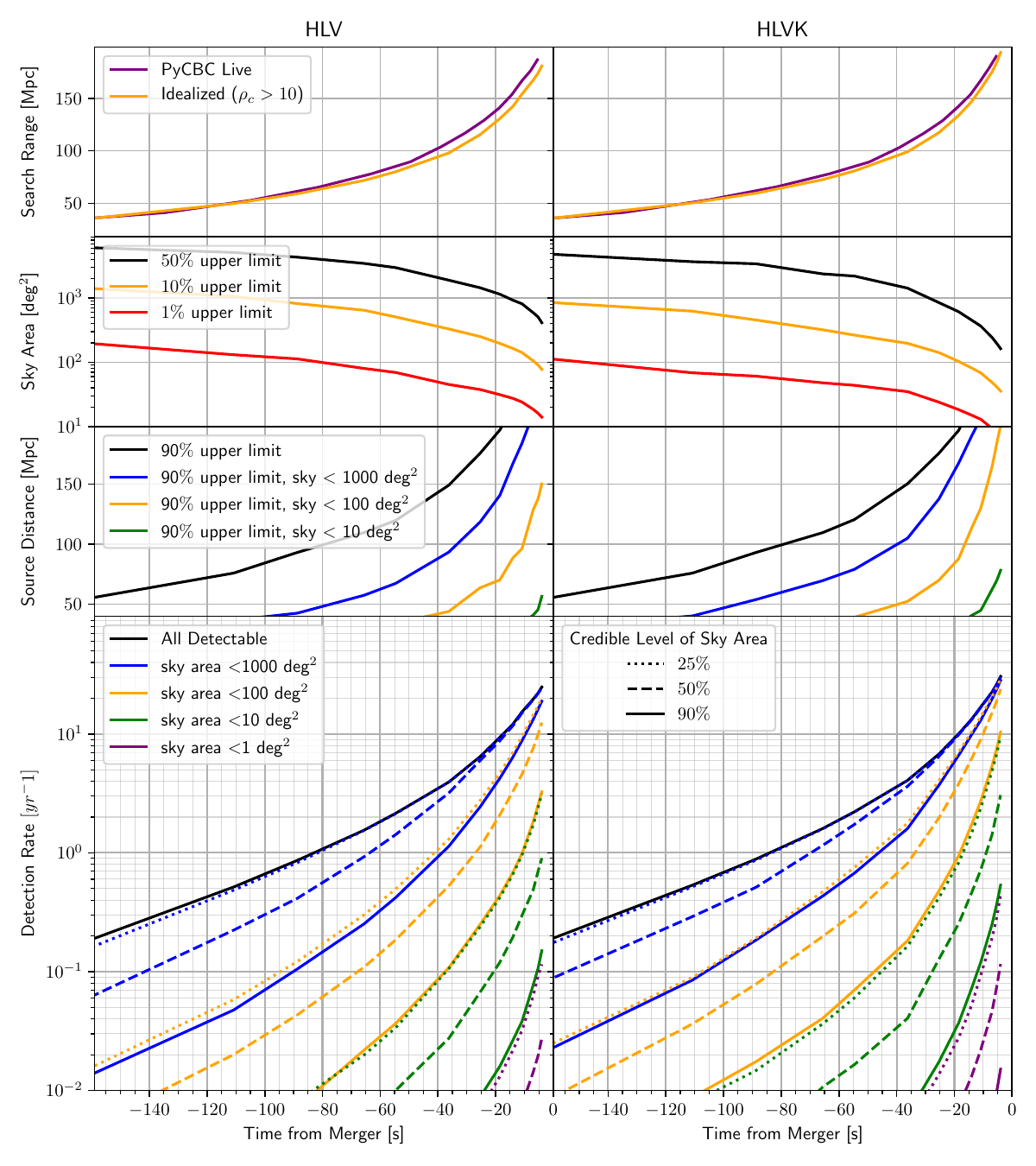}
    \caption{``Design'' era (2021-2022) detection and localization for the HLV network (left) and the full gravitational-wave detector network (right) as a function of time before merger for a fiducial 1.4-1.4$\msun$ BNS merger. (Top) The sky-averaged detection range for the idealized search and PyCBC Live operating at a false alarm rate of once per year. (Middle) The upper limit on the localization sky area and source distance, respectively, for detectable sources. Sky areas are quoted at the 90$\%$ credible level. (Bottom) The detection rate of all sources (black) and those that also have a sky localization less than 1000 deg$^2$ (blue), 100 deg$^2$ (orange), 10 deg$^2$ (green), or 1 deg$^2$ at a 90$\%$ (solid), 50$\%$ (dashed), and 25$\%$ credible level (dotted).}
\label{fig:design}
\end{figure*}

\begin{figure*}
    \centering
    \includegraphics[width=\textwidth]{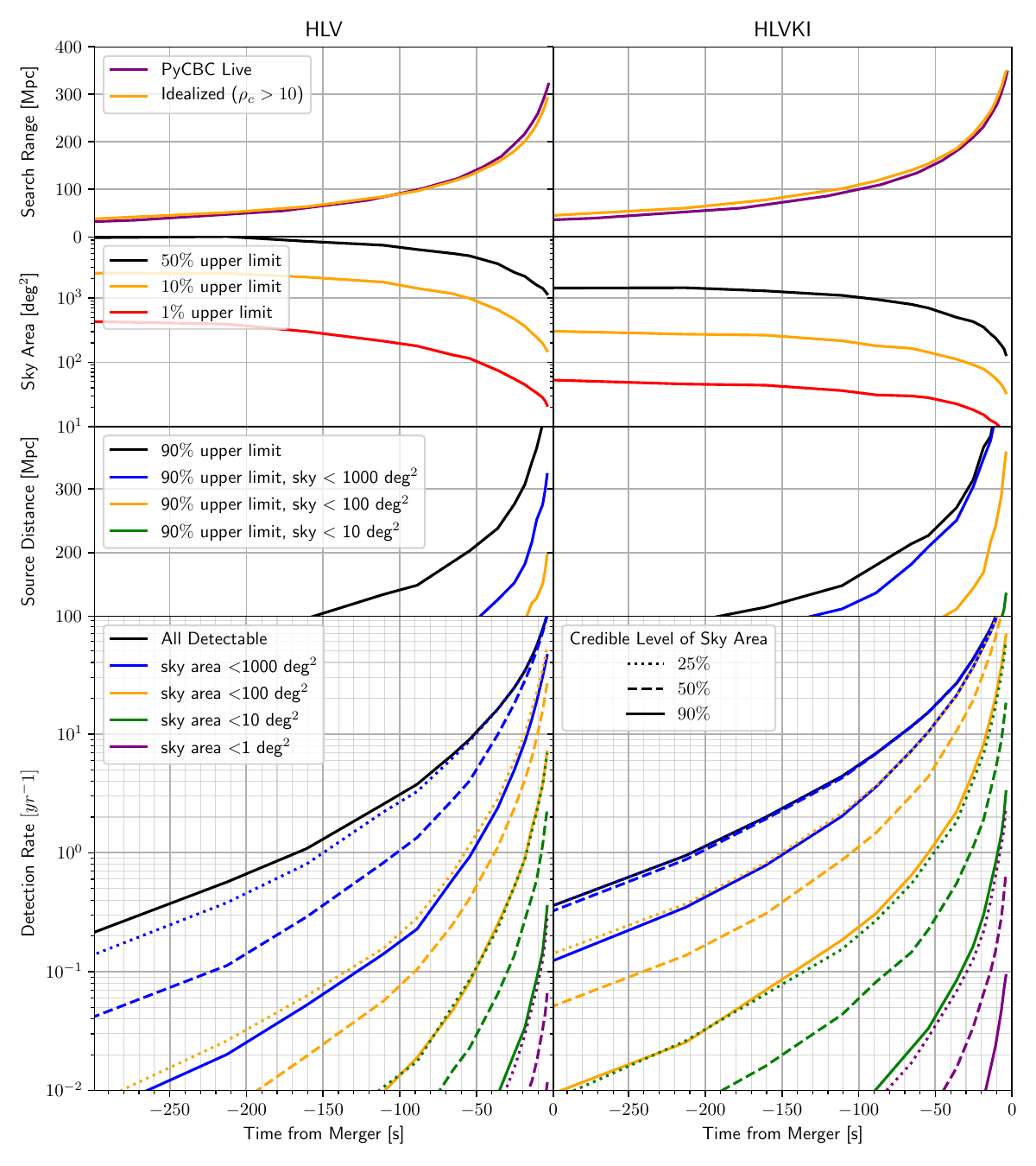}
    \caption{``A+'' era (2024-2026)  detection and localization for the HLV network (left) and the full gravitational-wave detector network (right) as a function of time before merger for a fiducial 1.4-1.4$\msun$ BNS merger. (Top) The sky-averaged detection range for the idealized search and PyCBC Live operating at a false alarm rate of once per year. (Middle) The upper limit on the localization sky area and source distance, respectively, for detectable sources. Sky areas are quoted at the 90$\%$ credible level. (Bottom) The detection rate of all sources (black) and those that also have a sky localization less than 1000 deg$^2$ (blue), 100 deg$^2$ (orange), 10 deg$^2$ (green), or 1 deg$^2$ at a 90$\%$ (solid), 50$\%$ (dashed), and 25$\%$ credible level (dotted).}
\label{fig:aplus}
\end{figure*}

\begin{figure*}
    \centering
    \includegraphics[width=\textwidth]{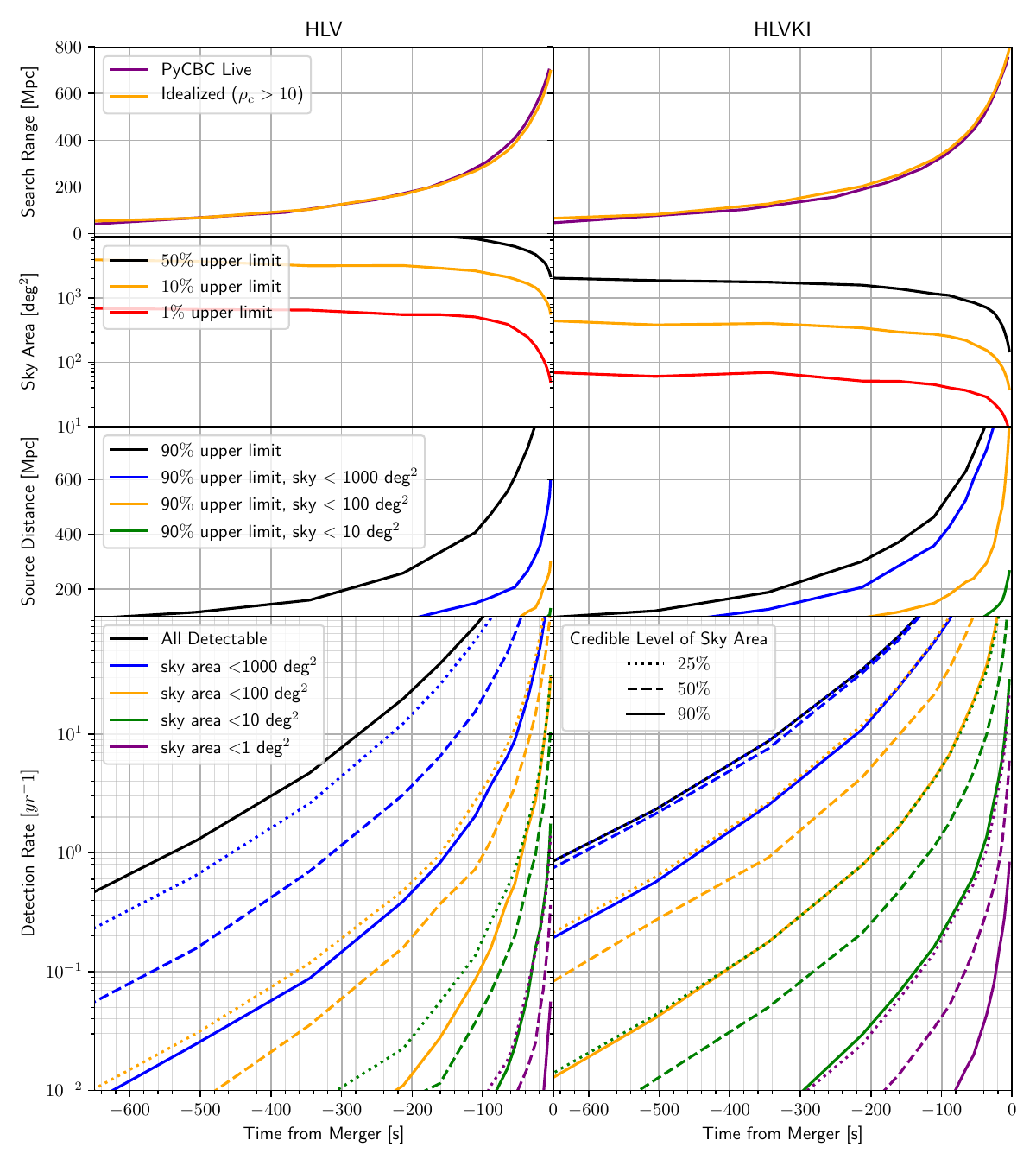}
    \caption{``Voyager'' era (late 2020's)  detection and localization for the HLV network (left) and the full gravitational-wave detector network (right) as a function of time before merger for a fiducial 1.4-1.4$\msun$ BNS merger. (Top) The sky-averaged detection range for the idealized search and PyCBC Live operating at a false alarm rate of once per year. (Middle) The upper limit on the localization sky area and source distance, respectively, for detectable sources. Sky areas are quoted at the 90$\%$ credible level. (Bottom) The detection rate of all sources (black) and those that also have a sky localization less than 1000 deg$^2$ (blue), 100 deg$^2$ (orange), 10 deg$^2$ (green), or 1 deg$^2$ at a 90$\%$ (solid), 50$\%$ (dashed), and 25$\%$ credible level (dotted).}
\label{fig:voyager}
\end{figure*}

Through the simulations described in Section \ref{sec:sims} we obtain the search sensitivity, expected rate of detections, and sky localization capabilities as a function of time before merger. These are shown in Figs.~\ref{fig:design},~\ref{fig:aplus}, and~\ref{fig:voyager} for the ``Design'', ``A+'', and ``Voyager'' era networks, respectively. For each era, we compare the reduced ``HLV'' detector network to the full network appropriate for that era. Note that the times shown in the horizontal axes do \emph{not} include the latency of the analysis, which we expect to vary over the years as technical improvements are made.

We find that the PyCBC Live low-latency search is already comparable to the idealized search, though some improvement may be possible for network configurations with a large number of detectors, whereas for three-detector configurations PyCBC Live already outperforms our simplified analysis when operating at false alarm rate of 1 per year. We can expect further improvements in pre-merger analyses to be made, but it is already clear that existing searches will be fully capable of meeting our predictions throughout the decade, assuming detector noise quality is comparable to previous observation runs.

\makeatletter\onecolumngrid@push\makeatother
\makeatletter\onecolumngrid@pop\makeatother

\makeatletter\onecolumngrid@push\makeatother
\makeatletter\onecolumngrid@pop\makeatother

\makeatletter\onecolumngrid@push\makeatother
\clearpage
\makeatletter\onecolumngrid@pop\makeatother

For $1.4-1.4 \msun$ BNS mergers and a merger rate density of $\sim1000 ~\mathrm{Gpc}^{-3}\mathrm{yr}^{-1}$~\citep{LIGOScientific:2018mvr,Abbott:2020uma}, we expect to detect one source per year with a $90\%$ credible sky localization area $<$ 100 deg$^2$ at $18$, $54$, or $195$~s before merger for the ``Design'', ``A+'', and ``Voyager'' full networks, respectively.  We note that detailed studies will be needed to determine the optimal follow-up strategy for specific observatories, however, we can explore some of the generic choices. Assuming a fixed observation sky area, facilities that would be willing to observe a higher rate of sources, and accept that the true source location may be outside the observed region a fraction of the time, will observe significantly more counterparts. For example, if an observatory targets every $50\%$ credible region with area less than 100 deg$^2$ the warning time is either increased to $34$, $104$, and $335$~s, respectively, or alternatively at the same warning times discussed earlier, we can expect $\sim 4-6$ observation opportunities per year instead of one. This more than doubles the expected number of observed counterparts given that half the time the true source location will be outside the observed region. Similarly, for ``A+'', and ``Voyager'', we find a single source per year will be localized $9$ and $43$~s before merger, respectively, with a $90\%$ credible sky area $<10~\mathrm{deg}^2$. For the $50\%$ credible region, we can increase the candidate rate to $\sim 6-8$ or increase the warning to $26$ and $115$~s, respectively.

If we restrict to only those sources within 100 (200) Mpc, we find that for the same fiducial rate of of BNS mergers, the ``Design'', ``A+'', and ``Voyager'' networks will detect and localize a single source per year with $90\%$ credible sky area $<$ 100 deg$^2$ with 17 (18), 54 (54), and 178 (194)~s of warning, respectively. There is little difference between the 100 and 200 Mpc cases as the vast majority of well localized sources will be at close distances. At 60s before merger, $90\%$ of the detected sources will be closer than 35 (13), 81 (28), and 231 (71) Mpc for the ``Design'', ``A+'', and ``Voyager'' networks, respectively, if we require that the source be localized with a $90\%$ credible area $<$ 100 (10) deg$^2$.

For systems detected at the earliest possible times considered, the sky localization typically evolves from an early multimodal distribution to a final unimodal or bimodal distribution, which is orders of magnitude more precise than the initial one. It is not uncommon for the initial localization to also be unimodal, though, but still orders of magnitude less precise than the final one. However, if we select cases where the initial localization is more precise than $\sim 100$ deg$^2$, we find most cases are already unimodal at the earliest time, i.e. they have a consistent overall direction throughout the inspiral.

\section{Application to other sources}
\label{sec:scale}

While for simplicity, we have reported results for a fiducial $1.4-1.4~\msun$ BNS merger, our results can be straightforwardly applied to other sources by scaling of the time and sensitive distance (or rate/volume as appropriate). The time axis scales inversely with the total mass of the source so that
\begin{equation}
    T_{m_1,m_2} = T_{1.4-1.4} \frac{2.8\msun}{m_1 + m_2}
\end{equation}
where $T_{1.4-1.4}$ is a time from our figures and $m_{1,2}$ are the desired source's redshifted component masses. This time rescaling directly accounts for the difference in localization for different mass sources, as the sky localization area is only dependent on the frequency bandwidth and the detector configuration for long duration signals, where the merger is above the detectors' sensitive frequency band.

The signal amplitude scales as the $\frac{5}{6}$th power of the source's chirp mass, which implies that volume and detection rate scale as 
\begin{equation}
    R_{m_1,m_2} = R_{1.4-1.4} \frac{(m_1 m_2)^{3/2}}{(m_1 + m_2)^{1/2}} \frac{2.8^{1/2}}{1.4^3}
\end{equation}
where $R_{1.4-1.4}$ is the rate of detections shown in our figures at a merger rate of 1000 Gpc$^{-3}$yr$^{-1}$. For illustrative purposes, if we assume that the rate of $1.4-4.0\msun$ sources were 100 Gpc$^{-3}$yr$^{-1}$ (which is consistent with limits reported in~\cite{LIGOScientific:2018mvr}), then we'd expect for the Voyager era to be able to have 70 seconds of warning for about one source per year with sky area $<100$ deg$^2$.

This same scaling may also be applied for heavy binary black hole mergers, as long as we only consider times before merger. After this time, the sky localization distribution is no longer accounted for by a simple time rescaling due to the signal terminating within the most sensitive frequency band.
For instance, considering GW190521~\citep{Abbott:2020tfl}, which may have merged within the accretion disk of a supermassive black hole and produced an optical counterpart~\citep{Graham:2020gwr}, we find that the time scale factor is $\sim 50$. Hence, even in an optimistic Voyager era, we could expect no more than a few seconds warning for similar mergers.

\section{Conclusions}

Achieving the goal of the prompt electromagnetic observation of a compact binary merger requires coordination across different observatories and cutting-edge instruments and facilities with wide fields of view, rapid pointing, and fully automated operation. By simulating a population of neutron star mergers, and the analysis of the associated data with current technology, we have shown that over the next decade the pre-merger warning time may increase by an order of magnitude from O(10) to O(100) seconds. For many telescopes, this will not yet be sufficient to re-point and tile a 100 deg$^2$ area~\citep{Coughlin:2019qkn}, although notable exceptions exist~\citep{Gehrels:2004aa,Sagiv:2013rma}, including Swift~\citep{Tohuvavohu:2020stm}, ZTF~\citep{Bellm_2018,Coughlin:2019xfb}, MASTER~\citep{2012ExA....33..173K}, and the CTA~\citep{Acharya:2013sxa}. Various facilities may also be able to use pre-merger warnings to alter triggering or observing configurations~\citep{James:2019xca}.

It is our hope that with the roadmap we provide, the observing community can plan for continued and automated operation of existing observatories, and envision bold new missions with varied observation bands and the goal of the first forecasted observation of a BNS merger within this decade. This includes concepts such as the Transient Astrophysics Probe~\citep{2019LPICo2135.5027C}. As GW170817 introduced gravitational waves to the field of multimessenger astronomy, we expect a multimessenger, multiband, prompt observation of a neutron star merger to be an important milestone in rapid time domain astronomy.

Data associated with the simulations is released at~\url{https://github.com/gwastro/gw-merger-forecasting}.

\acknowledgments
We thank Aaron Tohuvavohu, Eric Burns, Michael Coughlin and Nelson Christensen for their comments. This work was spurred by discussions and ideas at the Aspen Center for Physics, which is supported by National Science Foundation grant PHY-1607611. We acknowledge the Max Planck Gesellschaft and the Atlas cluster computing team at AEI Hannover for support.
\\
\\
Software: Numpy \citep{vanderWalt:2011bqk}, Scipy \citep{2020SciPy-NMeth}, Astropy \citep{Robitaille:2013mpa}, Matplotlib \citep{matplotlib}, ligo.skymap \citep{ligoskymap}, PyCBC \citep{pycbc-github}.

\bibliography{references}

\end{document}